\documentclass[prd,preprint]{revtex4}
\newcommand{\h}{\mathcal{H}}
\newcommand\R{\zeta}
\def\MM{M_{P}}

\newcommand\Lr{\rm L}
\newcommand\dr{\rm d}

\usepackage{xcolor}

\usepackage[utf8]{inputenc}

\usepackage{latexsym}
\usepackage{epsfig}
\usepackage{graphicx,subfigure}
\usepackage[normalem]{ulem}
\usepackage{color}
\usepackage{xcolor}
\usepackage{multirow}
\usepackage{hyperref}
\usepackage{numprint}
\usepackage{eucal}
\usepackage{amsmath}
\usepackage{amssymb}

\newcommand{\AER}[1]{{\color{black} #1}}

\newcommand{\AERR}[1]{{\color{black} #1}}
	
	\usepackage{amsmath}
	\usepackage{amsfonts}
  	\usepackage{amssymb}
	\usepackage{makeidx}
	\usepackage{amsfonts}
	\usepackage[ansinew]{}
	\usepackage[usenames,dvipsnames]{pstricks}
	\usepackage{epsfig}

    \usepackage{graphicx}

	\usepackage{float}

\newcommand{\be}{\begin{equation}}
\newcommand{\ee}{\end{equation}}
\newcommand{\bea}{\begin{eqnarray}}
\newcommand{\eea}{\end{eqnarray}}
\newcommand{\ba}{\begin{align}}
\newcommand{\ea}{\end{align}}

\makeindex

\begin{document}

\title{Constraining the effective field theory of dark energy with multimessenger astronomy}

\author{Antonio Enea Romano}

\affiliation{Instituto de Fisica,Universidad de Antioquia,A.A.1226, Medellin, Colombia}
\affiliation{ICRANet, Piazza della Repubblica 10, I--65122 Pescara, Italy}


\begin{abstract}

    The effective field theory  of dark energy predicts a possible time variation of the propagation speed of gravitational waves (GW) which could be tested with multimessenger astronomy.
    For this purpose we derive the relation between the redshift dependence of the propagation speed of GWs and the time delay between the detection of GWs and electromagnetic waves (EMWs) emitted by the same source.  According to the EFT the friction term of the GW propagation equation depends on the effective Planck mass and GW speed time variation, affecting the GW-EMW luminosity distance ratio. 
    
    We compute the general form of the GW-EMW luminosity distance ratio in terms of the effective GW speed and effective Planck mass, and then focus on  theories with constant Planck mass (CPM) and time varying GW speed. For CPM theories the GW speed can be jointly constrained by the GW-EMW detection time delay and luminosity distance ratio, allowing to derive a consistency relation between these two observables.  The event GW170817 and its EM counterpart satisfy the CPM consistency condition, and allows to set constraints on the time variation of the GWs speed,  and consequently on the coefficients of the effective theory. 
\end{abstract}

\keywords{}

\maketitle

\section{Introduction}
The detection of gravitational waves (GWs) \cite{LIGOScientific:2016aoc} by the Laser Interferometer Gravitational Wave Observatory (LIGO) and Virgo, has started the era of gravitational multi-messenger astronomy. 
GW events with an electromagnetic counterpart \cite{LIGOScientific:2017vwq}, also known as bright sirens, are used to test modified gravity theories (MGT) \cite{Baker:2017hug,Creminelli:2017sry,Sakstein:2017xjx,Ezquiaga:2017ekz,Wang:2017rpx} which predict a difference between the propagation speed of the GW and electromagnetic waves (EMW).
The event GW170817 set tight constraints on the GW-EMW speed difference, but these constraints were derived assuming a constant speed difference, while  the effective field theory approach \cite{Gleyzes:2013ooa,Creminelli:2014wna} predicts a possible time-dependence. Moreover, a frequency-dependence of the speed of GWs can arise in bigravity models, in Lorentz violating gravity models, as well as in theories of quantum gravity where the spectral dimension of spacetime changes with the probed scale. A general effective approach to study GWs was developed in \cite{Romano:2022jeh,Romano:2024apw,Romano:2023bzn}, showing that the speed of GWs could also depend on polarization.

In what follows,
we go beyond the constant GW-EMW difference approximation, and derive new relations for the time delay between the detection of GWs and EMWs emitted by the same source, as well as the redshift dependence of the GW-EMW speed difference.
For CPM theories the GW speed can  be jointly constrained by the GW-EMW time delay and the GW-EMW luminosity distance ratio, allowing to derive consistency relations between these two observables. 
The GW170817 observational data are analyzed, and the CPM consistency condition is shown to be satisfied. The data also allow to set constraints on the GWs speed, and a relation between the EFT action coefficients and the speed difference is derived, allowing to map the speed constraints into EFT coefficient constraints.

\section{Effective theory}
The quadratic effective field theory action (EFT) of perturbations for a single scalar  dark energy field was derived in \cite{Gleyzes:2013ooa} 

\be
\begin{split}
\label{total_action}
S = \int \! d^4x \sqrt{-g} \Bigg[ \frac{\MM^2}{2} f(t) R - \Lambda(t) - c(t) g^{00}  + \, \frac{M_2^4(t)}{2} (\delta g^{00})^2\, 
-\, \frac{m_3^3(t)}{2} \, \delta K \delta g^{00} -\, \\ m_4^2(t)\left(\delta K^2 - \delta K^\mu_{ \ \nu} \, \delta K^\nu_{ \ \mu} \right) + \frac{\tilde m_4^2(t)}{2} \, \R \, \delta g^{00}  \Bigg] \,,
\end{split}
\ee
where $\zeta$ is the curvature perturbation, $K_{\mu \nu}$ is the extrinsic curvature tensor, $\delta g^{00} \equiv g^{00} +1$, $\delta K_{\mu \nu} \equiv K_{\mu \nu} - H h_{\mu \nu}$, and $K \equiv K^{\mu}_{\ \mu}$ and $M_P$ is the Planck mass.
The above action for tensor modes gives 
\be
S_{\gamma}^{(2)} =\int d^4 x \, a^3 \frac{\MM^2 f }{8}\frac{1}{v^2_{\rm GW}} \left[   \dot{\gamma}_{ij}^2 -\frac{v^2_{\rm GW}}{a^2}(\partial_k \gamma_{ij})^2 \right]\,, \label{lgamma}
\ee
where the GWs speed is related to the EFT action coefficients by
\be
v_{\rm GW}^2 =  \left(1+\frac{2m_4^2}{\MM^2 f}\right)^{-1}\;.\label{vEFT}
\ee
In the literature of modified gravity the quantity $M^2_*=\MM^2 f /(8 \,v_{\rm GW}^2)$ is often introduced, in terms of which the action, using conformal time, takes the form
\be
S_{\gamma}^{(2)} =\int d^4 x \, a^2 M^2_* \left[   {\gamma'}_{ij}^2 -v^2_{\rm GW}(\partial_k \gamma_{ij})^2 \right]\,. \label{lgammaeta}
\ee
Note that $v_{\rm GW}$ depends on the ratio of two coefficients of the EFT action, $m_4$ and $f$, so that observational constraints on $v_{\rm GW}$ are  mapped into constraints of this ratio, not of the individual coefficients of the action. 


\section{Modified gravitational wave propagation}
It is convenient to rewrite the effective action in eq.(\ref{lgamma}) as
\be
S_{\gamma}^{(2)} =\int d^4 x \,  \frac{a^2 \Omega^2}{v^2_{\rm GW}} \left[   {\gamma'}_{ij}^2 -v^2_{\rm GW}(\partial_k \gamma_{ij})^2 \right]=\int d^4 x \,\alpha^2\Big[ \gamma_{ij}'^2-v_{\rm GW}^2 (\partial_k \gamma_{ij})^2\Big]~, \label{lgammaeta2}
\ee
where we have introduced $\Omega=M_* v_{ 
\rm GW}=\MM \sqrt{f/8}$, and we have defined 
\be
\alpha=\frac{a\, \Omega}{v_{\rm GW}}=a\, M_* \,.
\ee
\AERR{
In the EFT approach the coefficient of the Ricci scalar $\Omega$ is the quantity playing the role of effective Planck mass \cite{Kennedy:2017sof,Romano:2023ozy}, not $M_*$.
In modified gravity theories there exist alternative definitions of effective gravitational coupling $G_{eff}$, not based on the action as in the EFT, but on the perturbative equations, which are recast in the form of an effective modified Poisson's equation \cite{DeFelice:2011hq}
\be
\frac{k^2}{a^2}\Psi=-4\pi G_{eff}\delta\rho \,,
\ee
where $\Psi$ is one of the Bardeen's potentials and $\delta\rho$ is the energy density perturbation.
Recasting the perturbations equations in the above form implies that  $G_{eff}$ is not only dependent on $\Omega$, but it also includes other contributions from the additional fields related to the gravity modification. This can be convenient for studying the effects of the modification of gravity on structure formation, since it allows minimal modifications of the existing numerical codes, but it mixes in a single quantity different effects corresponding to different terms in the action, not just $\Omega$.}

\AERR{The  equation of motion corresponding to the effective action is }
\be
\gamma_{ij}''+2  \frac{\alpha'}{\alpha} \gamma_{ij}'-v^2_{\rm GW} \nabla^2 \gamma_{ij}=
\gamma_{ij}''+2  \h\Big(1-\frac{v_{\rm GW}'}{\h v_{\rm GW} }+\frac{\Omega'}{\h\Omega}\Big) \gamma_{ij}'-v^2_{\rm GW} \nabla^2 \gamma_{ij}=0  \,.\label{heft}
\ee

 After introducing  \cite{Romano:2023ozy} $\chi=\alpha\,\gamma_{ij}$ and taking the Fourier transform we obtain
\be
\chi_k''+\Big(v^2_{\rm GW} k^2 -\frac{\alpha''}{\alpha}\Big)\chi_k=0 \,. \label{chik}
\ee
We can write the solution  in the form $\chi_k(\eta)=A(\eta) e^{i B(\eta)}$ \cite{Nishizawa:2017nef} , which substituted in  eq.(\ref{chik}) gives 
\begin{align}
& \frac{2 A'}{A}+\frac{B''}{B'}=0 \,,   \label{hI}\\
&\frac{A''}{A}-\frac{\alpha ''}{\alpha }-B'^2+k^2 v_{\rm GW}^2\approx B'^2+k^2 v_{\rm GW}^2=0 \,,\label{hR}
\end{align}
where in the last equation we have applied the Wentzel–Kramers–Brillouin (WKB) approximation \cite{Nishizawa:2017nef,He:2022qcs}, consisting in neglecting the terms $A''/A$ and $\alpha''/\alpha$ w.r.t. $k^2 v_{\rm GW}^2$, since they vary on a cosmological time scale much larger than the GW period time scale.
Integrating the equations we get
\bea
B=\int^{\eta} \pm k\,v_{\rm GW}(\eta') d\eta' \,&,& A\propto \frac{1}{\sqrt{v_{\rm GW}}}\,,
\eea
which \AERR{imply} that the Fourier  mode $\gamma_{k,ij}$ of the GWs solution is
\bea
\gamma_{k,ij}&\propto&\frac{1}{\alpha\sqrt{v_{\rm GW}}} \exp{\left[i\int^{\eta} \pm k\,v_{\rm GW}(\eta') d\eta' \right]}=\frac{1}{\alpha\sqrt{v_{\rm GW}}}\exp{[\pm i k V(\eta)]} \,,
\eea
where we have defined the function $V(\eta)$ as
\be
V(\eta)=\int^{\eta} v_{\rm GW}(\eta') d\eta'\, ,\label{Veta}
\ee
which is related to the time delay w.r.t. the GR propagation time \cite{Nishizawa:2017nef}.
As an example, a general solution in physical space for a wave \AERR{propagating} along the $x$ direction, can be obtained in terms of $\gamma_k$ by using the inverse Fourier transform 
\be
\gamma_{ij}(\eta,x)=\frac{1}{\alpha\sqrt{v_{\rm GW}}} \frac{1}{2\pi}\int^{\infty}_{-\infty} d k f_k \exp{(i k x)} \exp{[\pm i k V(\eta)]}=\frac{1}{\alpha\sqrt{v_{\rm GW}}} f[x\pm V(\eta)] \,, \label{hx}
\ee
where $f_k$ denotes the Fourier transform of an arbitrary function $f$.
From eq.(\ref{hx}) we can see that the GW wave fronts propagate along  curves defined by $x\pm V(\eta)=const$, from which we get that along the GW propagation path
\be
dx = \frac{d V}{d\eta}d\eta=\mp v_{\rm GW}d\eta=\mp v_{\rm GW}\frac{dt}{a}\,.\label{dxMGT}
\ee
We show in  Appendix \ref{MGTS} that a similar result can be obtained for spherical waves, with an additional $1/r$ factor, giving
\bea
\gamma_{ij}=
&\propto& \frac{1}{r\,\alpha  \sqrt{v_{\rm GW}}}=\frac{\sqrt{v_{\rm GW}}}{r\,\Omega  } \,.
\eea
The above equation implies that the GW amplitude scales in time as
\be
\frac{\gamma_{ij,o}}{\gamma_{ij,e}}\propto \frac{\alpha_e}{\alpha_o}\sqrt{\frac{v_{\rm GW,e}}{v_{\rm GW,o}}}=\frac{a_e}{a_o}\frac{\Omega_e}{\Omega_o} \sqrt{\frac{v_{\rm GW,o}}{v_{\rm GW,e}}}\,, \label{hMGT}
\ee
where the subscripts $e$ and $o$ denote respectively  the emitted and observed quantities. This relation will be important to compute the effects of the modified GW propagation on the gravitational luminosity distance.
\section{Gravitational luminosity distance}
In general relativity (GR) the leading order calculation of the amplitude of GWs emitted by a binary coalescence gives 
\be
\gamma_{ij,e}\propto \frac{1}{r } {M_c}^{5/3} f_e^{2/3} \,, \label{hGR}
\ee
where $M_c$ is the chirp mass, and $f_e$ is the emitted GW frequency.
The  cosmological expansion implies an additional time dependence of the type
\be
\frac{\gamma_{ij,o}}{\gamma_{ij,e}}=\frac{a_e}{a_o} \label{ha}\,. 
\ee
Assuming matter to be minimally coupled to gravity, the redshift is defined as
\be
(1+z)=\frac{a_o}{a_e}\label{za}\,,
\ee
which allows to derive this useful relation
\be
\frac{1}{r}\frac{a_e}{a_o}=\frac{1}{r(1+z)}\frac{a_o}{a_o}=\frac{a_o}{d^{\rm EM}_{\Lr}}=\frac{1+z}{d^{\rm EM}_{\Lr}} \,,\label{aux}
\ee
where in the last equality we have set without loss of generality $a_e=1$,  and used the formula for the electromagnetic luminosity distance $d^{\rm EM}_{\Lr}$ in a flat Friedmann-Robertson-Walker (FRW) universe.
Due to the cosmological expansion, the emitted frequency $f_e$ is related to the observed frequency $f_o$ by 
\be
f_e=(1+z)f_o \,,
\ee
which combined with eq.(\ref{hGR}), eq.(\ref{ha}) and eq.(\ref{aux}) gives \cite{Finn:1992xs}
\be
\gamma_{ij,o}\propto \frac{1}{r}\frac{a_e}{a_o} {M_c}^{5/3} (1+z)^{2/3}f_o^{2/3}= \frac{1}{d^{\rm GW}_{\Lr}(z)} {M_c}^{5/3} (1+z)^{5/3}f_o^{2/3}=\frac{1}{d^{\rm GW}_{\Lr}(z)} {\mathcal{M}_c}^{5/3} f_o^{2/3} \,, \label{hGRo}
\ee
where we defined the redshifted chirp mass $\mathcal{M}_c=(1+z)M_c$  and the gravitational luminosity distance as
\be
d^{\rm GW}_{\Lr}(z)=a_o r(1+z)=d^{\rm EM}_{\Lr}(z) \,.
\ee
According to the above definition, in a flat FRW universe  the gravitational and electromagnetic luminosity distance are equal, if GWs propagate according to GR.

For GWs propagating according to the EFT equation(\ref{heft}) we can compute the gravitational luminosity distance using eq.(\ref{hMGT}), and following \AERR{a derivation similar to the one} given above for general relativity we obtain   
\be
d^{\rm GW}_{\Lr}(z)=d^{\rm EM}_{\Lr}(z) \frac{\Omega(0)}{\Omega(z)} \sqrt{\frac{v_{\rm GW}(z)}{v_{\rm GW}(0)}}=d^{\rm EM}_{\Lr}(z) \frac{M_*(0)}{M_*(z)} \sqrt{\frac{v_{\rm GW}(0)}{v_{\rm GW}(z)}}  \,.\label{rOmega}
\ee
The GW-EMW  luminosity distance ratio is affected by the ratio of the GW speed and of the effective Planck mass between the source and the observer. 
Expanding the distance ratio to first order in redshift we get
\be
\frac{d^{\rm GW}_{\Lr}(z)}{d^{\rm EM}_{\Lr}(z)}=1+  \left[\frac{ v_{\rm GW}'(0)}{2 v_{\rm GW}(0)}-\frac{\Omega '(0)}{\Omega (0)}\right] z =\AERR{1-\left[\frac{ v_{\rm GW}'(0)}{2 v_{\rm GW}(0)}+\frac{M_* '(0)}{M_* (0)}\right] z}\,,\label{rvomega}
\ee
showing that at low redshift the distance ratio depends on the relative derivative of the GW speed and of $\Omega$. 

The friction term is often re-written \cite{Belgacem:2018lbp} as 
\be
\frac{\alpha'}{\alpha}=\h(1-\delta)=\h\Big(1-\frac{v_{\rm GW}'}{\h v_{\rm GW} }+\frac{\Omega'}{\h\Omega}\Big)=\h\left(1+\frac{M_*'}{\h M_*}\right) \,,
\ee
from which after integration we obtain
\be
\frac{M_*(z)}{M_*(0)}=\exp{\left[\int^z_0\frac{\delta(z')}{1+z'}d\,z'\right]} \,,\label{mdelta}
\ee
\AERR{
implying
\be
\frac{M_*'(0)}{M_*(0)}=\frac{\Omega'(0)}{\Omega(0)}-\frac{v_{\rm GW}'(0)}{v_{\rm GW}(0)}=\delta(0).\label{mdelta0}
\ee
}
Using eq.(\ref{rvomega}) and eq.(\ref{mdelta0}), at first order in redshift the distance ratio can be approximated as
\be
\frac{d^{\rm GW}_{\Lr}(z)}{d^{\rm EM}_{\Lr}(z)}=1- \AERR{\left[\frac{ v_{\rm GW}'(0)}{2 v_{\rm GW}(0)}+\delta(0)\right]} z \,, \label{rdelta}
\ee
in agreement with eq.(50) in \cite{Belgacem:2018lbp} in the constant GW speed limit.

For constant GW speed (CGS) the friction term is related to the effective Planck mass \AERR{by}
\be
\frac{\Omega(z)}{\Omega(0)}=\exp{\left[\int^z_0\frac{\delta(z')}{1+z'}d\,z'\right]} \,,
\ee
while for constant Planck mass (CPM) theories, defined by $f=1$, it is related to the GW speed \AERR{by}
\be
\frac{v_{\rm GW}(z)}{v_{\rm GW}(0)}=-\exp{\left[\int^z_0\frac{\delta(z')}{1+z'}d\,z'\right]} \,.
\ee

As mentioned above, it is important to note that in the EFT the quantity $M_*=\Omega \, v_{\rm GW}$ is not playing the role of effective Planck mass \cite{Kennedy:2017sof}, which is instead given by $\Omega$, implying that the friction term depends both on the effective Planck mass and the speed, as shown in eq.(\ref{heft}). This distinction is not necessary for CGS gravity theories,  and in these theories \cite{LISACosmologyWorkingGroup:2019mwx} the modified friction parameter $\delta$ is only related to the effective Planck mass $\Omega$, denoted as $M_{eff}$ in \cite{LISACosmologyWorkingGroup:2019mwx}.

For CPM theories  eq.(\ref{rOmega}) reduces to
\be
r_{\dr}(z)=\frac{d^{\rm EM}_{\Lr}}{d^{\rm EM}_{\Lr}}=\sqrt{\frac{v_{\rm GW}(z)}{v_{\rm GW}(0)}}\,, \label{drEMC}
\ee
which allows to set constraints on the redshift evolution of the GW propagation speed using distance observations, and combine them with the constraints from the time delay. These joint constrains allow to derive a consistency relation to test the constancy of the Planck mass. 


\section{Constraints on GW-EMW propagation speed difference}

The relation between the time delay and the GW-EMW speed difference can be derived  by considering that the GW and EMW waves travel the same comoving distance between the source and the observer \cite{Jacob:2008bw,Mirshekari:2011yq,Kostelecky:2008be,Tason}.
As shown in Appendix \ref{MGTS}, for a spherical wave we have 
\be
dr=\mp v_{\rm GW}\frac{dt}{a} \,,
\ee
which can be integrated in redshift space using $dt=da/{\dot{a}}$ and the definition of redshift, giving 
\be
r_{\rm GW}(z)=\int^z \frac{v_{\rm GW}(z')}{H(z')} {\rm d} z' \,.
\ee
Assuming EMWs propagate along null geodesics, we obtain a similar relation 
\be
dr_{\rm EM} = v_{\rm EM} \, d\eta=\frac{v_{\rm EM}}{a}dt \,.
\ee

Since GWs and EMWs propagate between the same two points in space-time, they will travel the same comoving distance \cite{Jacob:2008bw}
\bea
r_{\rm GW}&=&\int^{t_{\rm s,\rm GW}}_{t_{\rm o,\rm GW}} \frac{v_{\rm GW}}{a(t')}{\rm d}t'=\int^{t_{\rm s,\rm EM}}_{t_{\rm o,\rm EM}} \frac{v_{\rm EM}}{a(t')}{\rm d}t'=r_{\rm EM} \,.\label{rcom}
\eea

Due to the difference in speed, in the above integrals we have distinguished between the observation times $t_{\rm o,\rm EM}$ and $t_{\rm o,\rm GW}$, while the emission times $t_{\rm s,\rm GW}$ and $t_{\rm s,\rm EM}$  could be different due to an emission time delay at the source. After a change of variable, we obtain the redshift space form \AERR{of} the integrals, which can be directly related to observations

\bea
r_{\rm GW}&=&\int^{z_{\rm s,\rm GW}}_{z_{\rm o,\rm GW}} \frac{v_{\rm GW}}{H(z')}{\rm d}z'=\int^{z_{\rm s,\rm EM}}_{z_{\rm o,\rm EM}} \frac{v_{\rm EM}}{H(z')}{\rm d}z'=r_{\rm EM} \,.\label{rcomt}
\eea
In the above equation we have distinguished between the GW and EMW redshift because if the corresponding speeds are different the consequent detection time delay would induce a difference in the redshift, and \AERR{also to} account for a possible emission time delay.

In what follows we adopt the notation $\Delta z_{\rm s}\equiv z_{\rm s,\rm GW}-z_{\rm s,\rm EM}$, $\Delta z_{\rm o}\equiv z_{\rm o,\rm GW}-z_{\rm o,\rm EM}$, and set $z_{\rm GW} \equiv z_{\rm s,\rm GW}$, $z_{\rm EM} \equiv z_{\rm s,\rm EM}$, $z_{\rm o,\rm EM}=0$.
Assuming photons propagate along null geodesics,  we can  express time  intervals in terms of redshift intervals using the equation
\bea
{\rm d}t&=&-\frac{{\rm d}z}{(1+z)H(z)} \label{dtdz} \,.
\eea

In the  EFT action in eq.(\ref{total_action}) the background dark energy corresponds to the term $\Lambda(t)$, and since observations are in good agreement with a flat $\Lambda CDM$ model, the low redshift evolution of $H(z)$ is well approximated by the corresponding low redshift Taylor expansion of the Friedman equation 
\be
H(z)\approx H_0(1+\frac{3}{2} \Omega_m z)\,. \label{Hz1}
\ee
Note that eq.(\ref{Hz1}) does not assume matter domination, and it corresponds to a constant $\Lambda(t)$, with $\Omega_{\Lambda}=1-\Omega_M$. Since the $\Lambda$CDM model is in good agreement with observational data, this should be a good low redshift effective approximation for $H(z)$ for a modified gravity theory able to fit the same observational data.
Substituting eq.(\ref{Hz1}) into eq.(\ref{dtdz}) we obtain

\bea
dt&\approx&-\frac{{\rm d}z}{(1+z_{\rm GW}) H_0(1+\frac{3}{2}\Omega_m z)} \,,\nonumber \\
\Delta t_{\rm o}&=&\frac{\Delta z_{\rm o}}{H_0} , \nonumber \\ 
\Delta t_{\rm s}&\approx&\frac{\Delta z_{\rm s}}{(1+z_{\rm GW})H_0 (1+\frac{3}{2} \Omega_m z_{\rm GW})}  \,,\label{dtdzos}
\eea
with the definitions $\Delta t_{\rm s}\equiv t_{{\rm s},{\rm EM}}-t_{{\rm s},{\rm GW}}$ and $\Delta t_{\rm o}\equiv t_{{\rm o},{\rm EM}}-t_{{\rm o},{\rm GW}}$, where  $\Delta t_{\rm s}$ is the emission time delay at the source,  while $\Delta t_{\rm o}$ is the observed time delay.

\AER{
\subsection{Varying GW speed with constant term}

Let us consider a time varying propagation speed of GWs, parametrized as 
\be
v_{\rm GW}=v_{\text{EM}}+\Delta v(z)=v_{\text{EM}}+\Delta v_0+\Delta v_1\,z \,,
\ee
with $\Delta v_0$ constant.
Using eq.(\ref{rcom}) and eq.(\ref{Hz1}) we can obtain the low redshift expansion of the comoving distances
\bea
r_{\rm GW}&=&\AERR{
\frac{(\Delta z_o-z_{\rm GW}) \Big[v_{\text{EM}} (3 \Delta z_o \Omega_m+3 \Omega_m z-4)+\Delta v_0 (3 \Delta z_o \Omega_m+3 \Omega_m z-4)-2 \Delta v_1 (\Delta z_o+z_{\rm GW})\Big]}{4 H_0}}\nonumber \\
&\approx& \frac{(z_{\rm GW}-\Delta z_o) \Big[2 \Delta v_0+2 v_{\text{EM}}+\Delta v_1 (\Delta z_o+z_{\rm GW})\Big]}{2 H_0} \,, \label{rgw}
\\
r_{\rm EM}&=&\frac{v_{\text{EM}} (z_{\rm GW}-\Delta z_{\rm s})}{H_0}\,. 
\label{2}
\eea
where to derive eq.(\ref{rgw}) we have used the fact that at low redshift $z\ll 1$, and assumed a small time delay, i.e. $\Delta z_o\ll 1$.

%
Equation (\ref{drEMC}) implies that for \AER{CPM theories} the observed GW-EMW distance ratio $r_d$ is related to the GW speed by 
\AER{
\be
r_{\dr}(z)=\frac{d^{\rm GW}_{\Lr}}{d^{\rm EM}_{\Lr}}=\left[ \frac{v_{\text{GW}}(z)}{v_{\text{GW}}(0)}\right]^{1/2}\approx 1 + \frac{1}{2}\frac{z \Delta v_1}{v_{\text{EM}} } \,. \label{rdv1}
\ee
}
In deriving the above equations we have used the low-redshift expansion of the EM luminosity distance 
\bea
d^{\rm EM}_{\Lr}(z_{\rm EM}) =(1+z_{\rm EM})\ r_{\rm EM}(z_{\rm EM}) \approx  (1+z_{\rm GW})\frac{v_{\rm EM}}{H_0}z_{\rm GW}~.\label{Dlow}
\eea
Note that the distance ratio $r_d$ can be different from $1$ only when the GW speed is not constant, and at low redshift the leading contribution is given by the linear term $\Delta v_1$.

Combining the condition $r_{\rm GW}=r_{\rm EM}$ with eq.(\ref{rdv1}) we obtain
\bea
\Delta v_0&=&\frac{(1+z_{\rm GW}) (\Delta z_o-\Delta z_s) v_{\text{EM}}^2}{d^{\rm EM}_{\Lr} H_0} +v_{\text{EM}}(1-r_d)\,, \label{dv0R}\\
\Delta v_1&=&\frac{2(r_d-1) (1+z_{\rm GW}) v_{\text{EM}}^2}{d^{\rm EM}_{\Lr} H_0} \,. \label{dv1R}
\eea
Rewriting the redshift intervals in terms of time intervals using \AERR{eq.(\ref{dtdzos})} we get an expression in terms of the observable time delay
\bea
\frac{\Delta v_0}{v_{\text{EM}}^2}&=&  \frac{1+z_{\rm GW}}{d_{\Lr}^ {\rm EM}}\Bigg[\Delta t_{\rm o}-(1+z_{\rm GW})\Delta t_{\rm s} \Bigg(1+\frac{3}{2} \Omega_m z_{\rm GW}\Bigg) \Bigg]+\frac{1}{ \, v_{\text{EM}}}(1-r_d) \,.\label{dv0Rt}
\eea

\subsection{Constant GW speed case}

Assuming $\Delta v_1=0$, from the condition $r_{\rm GW}=r_{\rm EM}$, solving for $\Delta v_0$  we get
\be
\Delta v_0=\frac{(\Delta z_{\rm o}-\Delta z_{\rm s}) v_{\text{EM}}}{z_{\rm GW}}\,, \label{dvz}
\ee
and it follows from eq.(\ref{dv1R}) that $r_d=1$, so that eq.(\ref{dv0Rt}) reduces to \cite{Tason}
\be 
\frac{\Delta v_0}{v_{\rm EM}^2}= \frac{1+z_{\rm GW}}{d_{\Lr}^ {\rm EM}}\Bigg[\Delta t_{\rm o}-(1+z_{\rm GW})\Delta t_{\rm s} \Bigg(1+\frac{3}{2} \Omega_m z_{\rm GW}\Bigg) \Bigg]\,.\label{dv0}\,
\ee


\subsection{Varying GW speed without constant term}

While this is just a special case corresponding to $\Delta v_0=0$,  it is useful to derive some relations which simplify the comparison with the constant case ($\Delta v_1=0$). 
From the condition $r_{\rm EM}=r_{\rm GW}$ we get

\be
\Delta v_1 \approx\frac{2(\Delta z_{\rm o}-\Delta z_{\rm s}) v_{\text{EM}}}{z_{\rm GW}^2}\,. \label{dv1z}
\ee
which using eq.(\ref{Dlow}) yields
\be
\Delta v_1 = \frac{2 (z_{\rm GW}+1)^2 (\Delta z_{\rm o}-\Delta z_{\rm s}) v_{\text{EM}}^3}{(d^{\rm EM}_{\Lr} H_0)^2}\,. \label{dv1dtz}
\ee
Using eq.(\ref{dtdzos}) we obtain the following expression for $\Delta v_1$ in terms of the time delay
\be
\begin{split}
\Delta v_1&=\frac{2 (z_{\rm GW}+1)^2 \ v_{\text{EM}}^3 }{{(d^{\rm EM}_{\Lr})}^2 H_0}  \times\Bigg[\Delta t_{\rm o}-
(1+z_{\rm GW})\Delta t_{\rm s}(1+ \frac{3}{2}\Omega_m z_{\rm GW})\Bigg]\,. \label{dv1dt}
\end{split}
\ee
 Note that eq.(\ref{dvz}) and eq.(\ref{dv1z}) imply $\Delta v_1=2 \Delta v/z_{\rm GW} $, a relation useful to check the constraints obtained from observations under different assumptions for the GW speed. The meaning of this relation is that the same comoving distance can be traveled by a GW traveling at speed $v_{\rm GW}=v_{\rm EM}+\Delta v_0$ or $v_{\rm GW}=v_{\rm EM}+\Delta v_1 z$, if $\Delta v_1=2 \Delta v_0/z_{\rm GW} $. Note in this case eq.(\ref{dv1R}) is also valid, i.e. the quantity $\Delta v_1$ can be determined from two independent observables.

\section{Consistency relation for constant Planck mass theories}
Assuming a constant Planck mass and a GW speed parametrized by $v_{\rm GW}=v_{\rm EM}+\Delta v_1 z$, there are two observables constraining the same quantity $\Delta v_1$, allowing to derive a consistency condition for CPM theories. 
Combining  the distance ratio (eq.(\ref{rdv1})) and time delay (eq.(\ref{dv1z})) relations, we obtain the CPM consistency relation (CR) 
\AER{
\be
\mathcal{E}(z_{\rm GW})\equiv r_{\dr}(z_{\rm GW})-\frac{(\Delta z_o-\Delta z_{\rm s})}{z_{\rm GW}} =1\,,  \label{CRz}
\ee
}
which using eq.(\ref{dtdzos}) can be expressed in terms of time delay intervals as
\AER{
\be
\begin{split}
\mathcal{E}(z_{\rm GW})&\equiv r_{\dr}(z_{\rm GW})-\frac{ (1 + z_{\rm GW})v_{\rm EM}}{d^{\rm EM}_{\Lr}}
\times\Bigg[\Delta t_{\rm o}- 
  (1+z_{\rm GW})\Delta t_{\rm s}\Bigg] \,=1\,. 
\label{CRt}
\end{split} 
\ee
}
For General Relativity $\Delta v_1=0$, implying  $r_{\dr}(z)=1$ and $\Delta t_{\rm o}=(1+z)\Delta t_{\rm s}$, which is consistent with the cosmological time dilation effect, and the CR is consequently satisfied.
A violation of the consistency relation would imply that, at least one of the two hypothesis used to derive it, is not correct, i.e. that the Planck mass is not constant, or that the GW speed has no redshift dependence.
If the CR is satisfied we can conclude that the Planck mass is not varying, and that observations are consistent with a redshift dependent GW speed or with GR.

\begin{table*}[t]
\centering
\begin{tabular}{ | m{1cm} | m{5.5cm}| m{5.5cm} | }
  \hline
   & $\Delta t_{\rm o}$ +  $d^{\rm GW}_{\Lr}$ & $\Delta t_{\rm o}$ +  $z_{\rm EM}$  \\ \hline
  $\Delta v_0$ & $-3 \times 10^{-15}  < \frac{\Delta v_0}{v_{\rm EM}}< 7 \times 10^{-16}$  & $-2 \times 10^{-15}  < \frac{\Delta v_0}{v_{\rm EM}}< 4 \times 10^{-16}$    \\ 
  \hline
\end{tabular}

\caption{Constraints on the GW speed from GW170817 and its electromagnetic counterpart observations, assuming a constant GW speed. The first column corresponds to the constraints from the observed time delay $\Delta t_{\rm o}$ and the gravitational luminosity distance, assuming $d^{\rm GW}_{\Lr}=d^{\rm EM}_{\Lr}$, as in \cite{LIGOScientific:2017zic} . The second column corresponds to the constraints from $\Delta t_{\rm o}$ and  the electromagnetic counterpart redshift $z_{\rm EM}$, computing $d^{\rm EM}_{\Lr}$ using the $\Lambda CDM$ model with the best fit parameters of \cite{Planck:2018vyg}.  Note that the constraint on \AERR{$\Delta v_0$} reported in the first column, obtained from $d^{\rm GW}_{\Lr}$ and $\Delta t_{\rm o}$, is the same as the one obtained in \cite{LIGOScientific:2017zic} without including the redshift effects, since at low redshift they are negligible. Constraints on the coefficients of the EFT can be  obtained from eq.(\ref{mfEFT}).}
\label{tabI}
\end{table*}

\section{Observational constraints from GW170817}

In this section we apply the theoretical results obtained above to constrain $\Delta v_1$ and \AERR{$\Delta v_0$} for different types of theories using the GW event GW170817 and its electromagnetic counterpart.

\subsection{Constant $v_{\rm GW}$ case}
Adopting an approach similar to \cite{LIGOScientific:2017zic}, we assume that the short-duration gamma-ray burst (GRB) signal was emitted 10 s after the GW signal, and use the observed time delay $(+1.74 \pm 0.05)$ s between the GRB 170817A and the GW170817 event \cite{Goldstein:2017mmi,GWOSC}. \AERR{Note that there exist also alternatives models \cite{LIGOScientific:2017zic} with longer source time delays $\Delta t_s$.}
Following the conservative approach adopted in \cite{LIGOScientific:2017zic}, we consider the lower bound of
the $90\%$ confidence interval of the gravitational wave luminosity distance $d^{\rm GW}_{\Lr}=26$ Mpc, assume $d^{\rm GW}_{\Lr}=d^{\rm EM}_{\Lr}$,  and hence determine the redshift $z_{\rm GW}=0.006$ using the $\Lambda$CDM cosmological parameters given by the Planck mission \cite{Planck:2018vyg}, obtaining \AERR{the} constraints shown in the first column of Table I. Note that the constraint on $\Delta v_0$ from $d^{\rm GW}_{\Lr}$ and $\Delta t_{\rm o}$ is the same as the one obtained in \cite{LIGOScientific:2017zic} without including the redshift effects, since at low redshift they are negligible.

Since optical follow-up observations have allowed to identify with high confidence the host galaxy as NGC 4993 \cite{Coulter:2017wya}, the redshift can be obtained  directly, and from it one gets the EM luminosity distance. Using $z_{\rm EM}=0.01$ and the same cosmological parameters we obtain the constraints listed in the second column of Table I. 
Note that the constraints using $z_{\rm EM}$ are better, since EM observations allow a high precision redshift measurement, and consequently a precise estimation of the EM luminosity distance  $d_L^{\rm EM}\approx 44$ Mpc, while the gravitational luminosity distance has a larger uncertainty, giving the smaller $90\%$ lower confidence value 26 Mpc.

\subsection{Constraints on varying $v_{\rm GW}$ without constant term}

\AER
{
For CPM theories there are two constraints on $\Delta v_1$, which can be combined together in the consistency condition for $\mathcal{E}(z)$ given in eq.(\ref{CRt}). Since for these theories the GW-EMW distance ratio can constrain independently the GW speed, we cannot assume  $d^{\rm GW}_{\Lr}=d^{\rm EM}_{\Lr}$ to derive the redshift, because this would imply $v_{\rm GW}(0)=v_{\rm GW}(z)$. In this case we need to use $z_{EM}$, and using eq.(\ref{dv1dt}) we obtain
\be
-4 \times 10^{-13}  < \frac{\Delta v_1}{v_{\rm EM}}< 8 \times 10^{-14}\,. \label{dv1C1}
\ee
From eq.(\ref{rdelta}), \AERR{eq(\ref{rvomega})} we obtain the following relations between the parameters of CGS and CPM theories fitting the same distance ratio observations 
\be
\frac{1}{2} \frac{\Delta v_1 }{v_{\text{EM}}}\Bigg|_{CPM}=-\frac{\Omega'(0)}{\Omega(0)}\Bigg|_{CGS}=-\delta(0)\Bigg|_{CGS}  \,,
\ee
\AERR{which is also consistent with eq.(\ref{mdelta0}) in the constant GW speed limit.}
}

\subsection{Varying GW speed with constant term}
In this case we need two observables, the time delay and the distance ratio to obtain an algebraic expression for the two parameters $\Delta v_0,\Delta v_1$.
The GW170817  90\% confidence bounds on the two parameters are
\bea
-88.1< \frac{\Delta v_1}{v_{\text{EM}}}< 10.5 \,, \\
-0.05<\frac{\Delta v_0}{v_{\text{EM}}}<0.44 \,.
\eea
The bound on $\Delta v_1$ is weaker than the one in eq.(\ref{dv1C1}) because in this case $\Delta v_1$ is  determined by the distance ratio $r_d$ in eq.(\ref{dv1R}), which has a large error. When we assume $\Delta v_0=0$ the quantity $\Delta v_1$ can be determined from the time delay, using eq.(\ref{dv1dt}), or from the distance ratio, using eq.(\ref{dv1R}), and the constraint in eq.(\ref{dv1C1}) is the one obtained form time delay observations, which have smaller error compared to the GW distance, implying a tighter constraint.

Mathematically this can be understood from the fact that we have two relations, one for the time delay, the other for the distance ratio, and when we assume $\Delta v_0=0$ they both give a constraint on $\Delta v_1$, with the one from the time delay being the strongest, as reported in eq.(\ref{dv1C1}). When $\Delta v_0$ is not zero, we need two relations to solve for the  quantities \{$\Delta v_0,\Delta v_1$\}, and consequently they are less constrained. On the contrary when we try to constrain only one parameter, assuming the other to be zero, we can use the most precise observation, which is the time delay, and obtain stronger constraints.
}

\subsection{Testing the CPM consistency condition}
Assuming the GW event to have been hosted by NGC 4993 we  can obtain $d^{\rm EM}_{\Lr}$ from it, while for $d^{\rm GW}_{\Lr}$ we take the $90\%$ confidence interval, obtaining
\be
0.56<\mathcal{E}(z)<1.05 \,.
\ee
\AER{The constraint is consistent with a constant Planck mass.}
The constraints on the EFT coefficients can be derived from eq.(\ref{vEFT}), which in the small $\Delta v$ limit gives
\be
\frac{m_4^2}{\MM^2 f}=-\frac{\Delta v(z)}{v_{\rm EM}}\,. \label{mfEFT}
\ee
For CPM theories, for which $f=1$, the above equation gives a direct relation between the EFT coefficient $m_4$ and $\Delta v$, while for \AER{non CPM} theories only the ratio between $m_4$ and $f$ is constrained. The combination with other observational data sets can resolve this degeneracy, and we leave this to a future work.

\section{Conclusions and Discussion}
The EFT predicts that GWs speed could be different from the speed of light, and time dependent. Since a single multimessenger event has been observed so far, we cannot exclude that this time dependency will be observed with higher redshift observations. 
The expected improved sensitivity of future GW detectors will allow the observation of an increasing number of bright sirens, leading to model independent constraints on the time variation of the speed of GWs.
Using the observations of the event GW170817 and its EM counterpart we have obtained  constraints on the variation of the speed of GWs from the speed of light for frequencies at the kHz range.
An independent way to test modified gravity models is provided through the \AER{CPM consistency condition}, eq.(\ref{CRt}), and we found that the event GW170817 does not violate the consistency condition, i.e. there is no evidence of a varying  Planck mass.
Finally, let us note that in  deriving of  eq.(\ref{dv1dt}), eq.(\ref{dv0})  and eq.(\ref{CRt}) we have made some low-redshift approximations, which may not be justified for higher redshift observations. In this case higher order redshift expansions can be used, 
or eq.(\ref{rcom}) can be integrated numerically.

Gravitational waves observation do no allow to constrain individually the coefficients of the EFT action, so in the future it will be important to combine different observational data sets to resolve this degeneracy.
We have used effective field theory to obtain model independent relations and constraints, but in the future it will be interesting to map these into constraints on specific modified gravity theories.
\section{Acknowledgments}
This material is based upon work supported by NSF's LIGO Laboratory which is a major facility fully funded by the National Science Foundation.
We thank Sakellariadou, Nathan K. Johnson-McDaniel, Tessa Baker, Jay Tasson,  Nicola Tamanini and Sergio Vallejo for useful comments and discussions. AER was supported by the UDEA projects 2021-44670, 2019-28270, 2023-63330.

\appendix

\section{Modified spherical waves propagation}
\label{MGTS}
For a spherically symmetric wave $\gamma_{ij}(\eta,r)$ in spherical coordinates the GW propagation eq.(\ref{heft}) takes the form
\be
\gamma_{ij}''+2  \frac{\alpha'}{\alpha} \gamma_{ij}'-v^2_{\rm GW} \frac{1}{r^2}\frac{\partial}{\partial r^2}(r^2 \gamma_{ij})=0~. \label{heftr}
\ee
Defining $\chi=\alpha \,r \, \gamma_{ij}$ and taking the Fourier transform w.r.t. to the radial coordinate $r$ we get 
\be
\chi_k''+\Big(v_{\rm GW} k^2 -\frac{\alpha''}{\alpha}\Big)\chi_k=0 \,. \label{chirk}
\ee
The above equation has the same form of eq.(\ref{chik}), implying it can be solved with same the WKB approximation used to solve eq.(\ref{chik}), and after taking the inverse Fourier transform with respect to the radial coordinate $r$ we obtain a general solution of the form
\be
\gamma_{ij}(\eta,r)=\frac{1}{\,r \alpha\sqrt{v_{\rm GW}}} \frac{1}{2\pi}\int^{\infty}_{-\infty}  d k f_k \exp{(i k r)} \exp{[\pm i k V(\eta)]}=\frac{1}{\,r \alpha \sqrt{v_{\rm GW}}} f[r\pm V(\eta)] \,, \label{hr}
\ee
where $f$ is an arbitrary function,  $f_k$ denotes its Fourier transform, and $V(\eta)$ is defined in eq.(\ref{Veta}).
Wave fronts propagate along curves defined by $r\pm V(\eta)=const$,
from which we get that
\be
dr = \frac{d V}{d\eta}d\eta=\mp v_{\rm GW}d\eta=\mp v_{\rm GW}\frac{dt}{a} \,.\label{drMGT}
\ee
 
\bibliographystyle{h-physrev4}
\bibliography{mybib}
\end{document}